\documentclass[usenatbib]{mnras}
\usepackage{newtxtext,newtxmath}

\usepackage[T1]{fontenc}
\usepackage{ae,aecompl}

\usepackage{graphicx}
\usepackage{color} 
\usepackage{pbox} 

\definecolor{gray}{RGB}{150,150,150}

\newcommand{\degree}{$^{\circ}~$}

\title[A cometary surface evolution model]{Constraints on cometary surface evolution derived from a statistical analysis of 67P's topography}
\author[J.-B. Vincent et al.]{
\parbox{\textwidth}{J.-B. Vincent$^{1}$\thanks{E-mail: \texttt{jean-baptiste.vincent@dlr.de}}
S.F. Hviid$^{1}$
S. Mottola$^{1}$
E. Kuehrt$^{1}$
F. Preusker$^{1}$
F. Scholten$^{1}$
H. U. Keller$^{2}$
N. Oklay$^{1}$
D. de Niem$^{1}$
B. Davidsson$^{3}$
M. Fulle$^{4}$
M. Pajola$^{5,6}$
M. Hofmann$^{7}$
X. Hu$^{7}$
H. Rickman$^{8, 9}$
Z.-Y. Lin$^{29}$
C. Feller$^{11}$
A. Gicquel$^{3}$
S. Boudreault$^{7}$
H. Sierks$^{7}$
C. Barbieri$^{6}$
P. L. Lamy$^{13}$
R. Rodrigo$^{14, 15}$
D. Koschny$^{16}$
M. F. A'Hearn$^{10}$
M. A. Barucci$^{11}$
J.-L. Bertaux$^{17}$
I. Bertini$^{6}$
G. Cremonese$^{18}$
V. Da Deppo$^{19}$
S. Debei$^{20}$
M. De Cecco$^{21}$
J. Deller$^{7}$
S. Fornasier$^{11}$ 
O. Groussin$^{13}$
P. J. Guti\'errez$^{22}$
P. Guti\'errez-Marquez$^{7}$ 
C. G\"uttler$^{7}$ 
W.-H. Ip$^{23, 29}$ 
L. Jorda$^{13}$ 
J. Knollenberg$^{1}$
G. Kovacs$^{7, 30}$
J.-R. Kramm$^{7}$
M. K\"uppers$^{25}$
L. M. Lara$^{22}$
M. Lazzarin$^{6}$
J. J. Lopez Moreno$^{14}$
F. Marzari$^{12}$
G. Naletto$^{24, 19, 6}$ 
L. Penasa$^{26}$
X. Shi$^{7}$
N. Thomas$^{27}$
I. Toth$^{28, 13}$
C. Tubiana$^{7}$}
\\~\\
\parbox{\textwidth}{
$^{1}$Institute of Planetary Research, DLR, Rutherfordstrasse 2, D-12489 Berlin,Germany;
$^{2}$Institute for Geophysics and Extraterrestrial Physics, TU Braunschweig,D-38106 Braunschweig, Germany;
$^{3}$NASA Jet Propulsion Laboratory, 4800 Oak Grove Drive, Pasadena, CA91109, USA;
$^{4}$INAF – Osservatorio Astronomico di Trieste, via Tiepolo 11, I-34143 Trieste,Italy;
$^{5}$NASA Ames Research Center, Moffett Field, CA 94035, USA;
$^{6}$Centro di Ateneo di Studied Attivit\`a Spaziali 'Giuseppe Colombo' (CISAS), University of Padova, Via Venezia 15, I-35131 Padova, Italy;
$^{7}$Max-Planck Institut f\"ur Sonnensystemforschung, Justus-von-Liebig-Weg, 3 D-37077 Goettingen, Germany;
$^{8}$PAS Space Research Center, Bartycka 18A, PL-00716 Warszawa, Poland;
$^{9}$Department of Physics and Astronomy, Uppsala University, Box 516, SE-75120 Uppsala, Sweden;
$^{10}$Department for Astronomy, University of Maryland, College Park, MD20742-2421, USA;
$^{11}$LESIA, Observatoire de Paris, CNRS, UPMC Univ Paris 06, Univ. Paris-Diderot, 5 Place J. Janssen, F-92195 Meudon Pricipal Cedex, France;
$^{12}$Department of Physics and Astronomy 'G. Galilei', University of Padova,Vic. Osservatorio 3, I-35122 Padova, Italy;
$^{13}$Aix Marseille Universit\'e, CNRS, LAM (Laboratoire d'Astrophysique deMarseille) UMR 7326, F-13388 Marseille, France;
$^{14}$Centro de Astrobiologia (INTA-CSIC), European Space Agency (ESA),European Space Astronomy Centre (ESAC), PO Box 78, E-28691 Villanuevade la Canada, Madrid, Spain;
$^{15}$International Space Science Institute, Hallerstrasse 6, CH-3012 Bern,Switzerland;
$^{16}$Scientific Support Office, European Space Agency, NL-2201 Noordwijk,the Netherlands;
$^{17}$LATMOS, CNRS/UVSQ/IPSL, 11 Boulevard d’Alembert, F-78280 Guyancourt,France;
$^{18}$INAF Osservatorio Astronomico di Padova, Vicolo dell'Osservatorio 5,I-35122 Padova, Italy;
$^{19}$CNR-IFN UOS Padova LUXOR, Via Trasea 7, I-35131 Padova, Italy;
$^{20}$Department of Industrial Engineering University of Padova Via Venezia,1, I-35131 Padova, Italy;
$^{21}$University of Trento, via Sommarive, 9, I-38123 Trento, Italy;
$^{22}$Instituto de Astrofisica de Andalucia-CSIC, Glorieta de la Astronomia,E-18008 Granada, Spain;
$^{23}$Space Science Institute, Macau University of Science and Technology,Taipa, Macau;
$^{24}$Department of Information Engineering, University of Padova, ViaGradenigo 6/B, I-35131 Padova, Italy;
$^{25}$ESA/ESAC, Camino Bajo del Castillo s/n, Ur. Villafranca del Castillo,E-28692 Villanueva de la Canada, Madrid, Spain;
$^{26}$Dipartimiento di Geoscienze, University of Padova, via Granedigo 6, I-35131 Padova, Italy;
$^{27}$Physikalisches Institut, Sidlerstrasse 5, University of Bern,CH-3012 Bern,Switzerland;
$^{28}$Observatory of the Hungarian Academy of Sciences, PO Box 67, 1525Budapest, Hungary;
$^{29}$Institute of Astronomy, National Central University, 32054 Chung-Li, Taiwan;
$^{30}$Budapest University of Technology and Economics,Department of Mechatronics, Optics and Engineering Informatics, Muegyetem rkp 3, 1111 Budapest, Hungary}
}

\date{Accepted 2017 June 29. Received 2017 June 28; in original form 2017 March 24}

\pubyear{2017}

\begin{document}
\label{firstpage}
\pagerange{\pageref{firstpage}--\pageref{lastpage}}
\maketitle

\begin{abstract}
We present a statistical analysis of the distribution of large scale topographic features on comet 67P/Churyumov-Gerasimenko. We observe that the cumulative cliff height distribution across the surface follows a power law with a slope equal to $-1.69 \pm 0.02$. When this distribution is studied independently for each region, we find a good correlation between the slope of the power law and the orbital erosion rate of the surface. For instance, the northern hemisphere topography is dominated by structures on the 100~m scale while the southern hemisphere topography, illuminated at perihelion, is dominated by 10~m scale terrain features.
Our study suggest that the current size of a cliff is controlled not only by material cohesion but by the dominant erosional process in each region. This observation can be generalized to other comets, where we argue that primitive nuclei are characterized by the presence of large cliffs with a cumulative height power index equal to or above -1.5, while older, eroded cometary surfaces have a power index equal to or below -2.3.
In effect, our model shows that a measure of the topography provides a quantitative assessment of a comet's erosional history, i.e. its evolutionary age.
\end{abstract}

   \begin{keywords}
 comets: general, comets: individual: 67P
  \end{keywords}
  \newpage~
%
\section{Introduction}
One of the many surprises revealed by ESA's Rosetta spacecraft was the complex landscape of comet 67P/Churyumov-Gerasimenko. The surface is rich in land forms comparable to what is usually found on larger planetary bodies. Rosetta has mapped extensively the comet and its morphology has been described in many publications: \cite{thomas2015, elmaarry2015a, elmaarry2016, giacomini2016, birch2017}. Among all morphological features, we focus our interest on the near-vertical walls of cliffs and pits, interpreted to result from of surface collapse \citep{vincent2015b} and which clearly display ongoing regressive erosion due to ongoing activity/thermal stress \citep{vincent2016a} or sudden outbursts \citep{vincent2016b}.

Cliffs on 67P/Churyumov-Gerasimenko were not unexpected. Similar features have been observed previously on most other nuclei visited by spacecraft: 19P/Borelly \citep{britt2004}, 81P/Wild 2 \citep{brownlee2004}, 9P/Tempel 1 \citep{thomas07} but Rosetta provided the opportunity to look at these features in greater detail, and for an extensive period of time. We could for instance characterize the boulder size distribution in the cliffs' taluses \citep{pajola2015, pajola2016a}, and link observed collapses to activity \citep{vincent2016b, pajola2017}. 

In this work, we investigate the relation between cliffs or other vertical features and the erosional rates and material strengths. While we do not understand yet how cliffs are formed on a comet, the simple fact that they exist puts constrains on the material strength. Indeed, even in a very low gravity environment (typically $2.10^{-4} m.s^{-1}$, see section \ref{sec:gravity}), cliffs without strength would naturally collapse under their own weight in a few minutes \citep{jeffreys1952}. As cliffs were clearly stable for at least the two years time span of the Rosetta mission, the material properties must be sufficient to ensure their existence.

The surface strength on comet 67P has been investigated in localized areas and values published in several papers. For instance \citep{vincent2015b} constrained the strength of material surrounding active pits, interpreted as sink holes; \cite{groussin2015a} measured the strength of stable overhangs in selected areas of the comet; \cite{biele2015} and \cite{spohn2015} computed local strength respectively from the Philae lander bounce on Agilkia, and the MUPUS measurements at Abydos. All authors agree on a typical tensile strength in the range 10-100 Pa, and a compressive strength in the kPa range for the dusty layer, up to a couple of MPa for the underlying consolidated material.

While these different studies are converging, their scope was limited to very specific regions of the comet and may not fully describe the material. Additionally, strength alone may not be the main driver for the topography, as evolutionary processes can play a significant role. Therefore, our aim is to derive global statistics on the topography across the entire surface of 67P and link this to our current understanding of material strength and the variable evolutionary history of the nucleus.

\section{Data and methods}

\subsection{Shape model}
Our analysis is based on the most accurate 3-dimensional reconstruction of 67P's nucleus topography, obtained by photogrammetry. The data set and technique are described in \cite{preusker2015} for the Northern hemisphere. This paper uses a new version of the 3d shape ("cg-dlr\_spg-shap7-v1.0\_500Kfacets.ply"), representing the complete nucleus, and presented in \cite{preusker2017}. The full resolution model comprises about 22 million vertices arranged in 44 million triangular facets. Vertex positions have a typical spacing of 1-2~m and 1-sigma accuracy of 0.2-0.3~m. The typical uncertainty in the facet orientation is in the order of 2-5\degree.

Processing such a large data set is computationally prohibitive, while the full resolution is not necessary for our analysis the typical feature size is larger than 10 meters. We therefore based this study on a decimated version of the same shape model, with about 250 000 vertices and 500 000 facets. On average, vertices are separated by a distance of about 15~m.

\subsection{Gravity}\label{sec:gravity}
In order to define which structures are actually cliffs, we need first to estimate the surface effective gravity, the combination of gravitational acceleration and centrifugal force due to the rotation. On a body such as 67P ($1.5~km$ mean radius, $1 \times 10^{13}~kg$ mass, 12.4 hours rotation period), the mean gravity is in the order of $2 \times 10^{-4} m.s^{-2}$ and the centrifugal force is about $3.10^{-5}~m.s^{-2}$. Hence, the centrifugal force opposes gravity with a relative magnitude of up to 15\% and must be accounted for in our calculation.

Gravity values are obtained for each facet using the classical \cite{werner1996} approach. Because gravity calculation on a convex body is non-trivial, we also compared our results with an alternative model by \cite{cheng2012}. For the 500k facets shape, the absolute difference between the two gravity models is in the range [$1.9 \times 10^{-7} to 3.7 \times 10^{-6}~m.s^{-2}$], i.e. less than 1\% of the effective gravity. As both methods use an independent approach, we are therefore convinced that we have calculated a reliable approximation of the gravity vector on each facet.\\

We note that using a simpler model (two central masses and the ellipsoid parameters described in \cite{jorda2016}) is not sufficient. While the gravity obtained agrees with the more advanced models for most of the surface, we found that the simple model leads to anomalously large gravity values (greater than twice the expected figure) for about 8\% of the facets, especially in highly concave areas, such as the Hapi/Hathor region.

\subsection{Slopes and automatic detection of cliffs}
We also measure the effective surface slope, defined as the angle between the surface normal vector and the opposite of the local gravity. A slope of 0\degree is flat with respect to gravity, while a slope of 90\degree describes a cliff.

\begin{figure*}
   \centering
   \includegraphics[width=0.49\hsize]{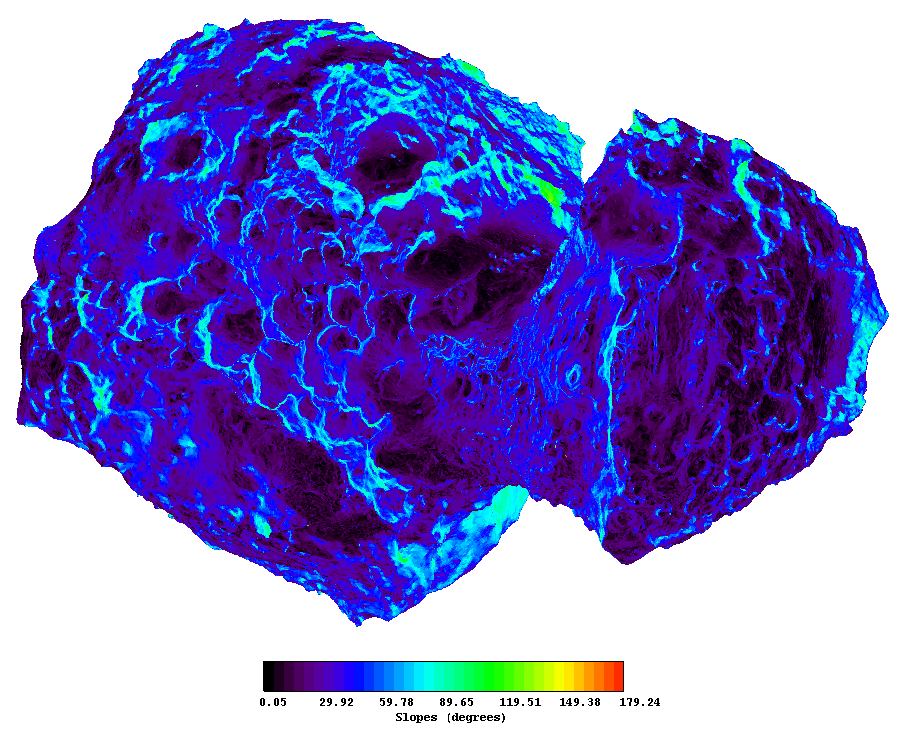}
   \includegraphics[width=0.49\hsize]{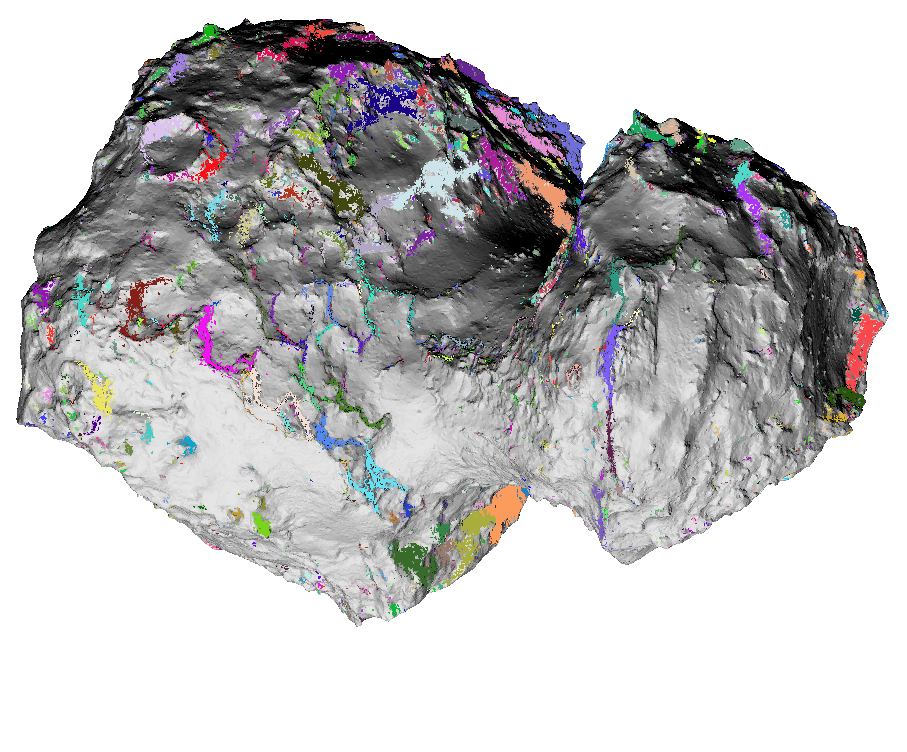}
   \caption{Visual representation of the automatic detection of topographic features of interest. Left panel shows the effective gravitational slope for each facet of the shape model (accounting for gravity + centrifugal force). Right panels shows cliffs, i.e. independent sets of connected facets with a slope larger than 60\degree. Colors indicate different cliffs.}
    \label{fig:cliffs_3d}
\end{figure*}

Using this measure of the slope as input, we developed an algorithm to automate the detection of all topographic features relevant to this study. It works in three consecutive steps:
\begin{enumerate}
\item We isolate facets of the shape model having a slope larger than 60\degree. This arbitrary value is taken as a very conservative maximum angle of repose on 67P. It is twice the value measured for granular material in granular flows observed in various regions of the comet (30\degree, \cite{vincent2016a}). Using a high angle ensures that none of the selected areas contain loose dust. Additionally, it prevents us from selecting artefacts. Indeed, by reducing the number of facets, the decimation process from 44 million to 500 thousand facets smooths out features with a size comparable to the facet length. For instance, a boulder of 15 m height may end up being described with one vertex only and show lateral slopes close to 45\degree. The choice of this slope angle limit effectively defines the lowest height that can be detected: \mbox{height $>$ min(length) $\times$ sin(slope) $=~13~m$.}
\item We then grouped together neighbouring high slopes by geographic location. Starting with the facet identified in step one as having a slope $>$60\degree, we then find all the neighbouring facets that match that criterion and group them into a unique set. We then iterate over these newly added facets until there are no more remaining neighbours with $>$60\degree slopes to be added to the current set. We select another cliff not yet in a set and repeat the process. Thus, we end up with a separation of all cliffs as independent entities with no feature being identified twice. Figure \ref{fig:cliffs_3d} shows a 3d visualization of the identified cliffs. Our algorithm properly separates features that belong to the same morphological region. For instance the inner walls of a pit are grouped together, while the facets surrounding a large outcrop are similarly grouped.
\item For each topographic feature identified in this way, we extract and save parameters that can be used for further investigation: average 3d position on the shape model, local gravity, slope, height, area. The height is defined relative to the local gravity: We first project the three-dimensional positions of all vertices in a set (that is all facets describing a feature) onto the local gravity vector. We then define the height as the altitude difference between the highest and lowest point of the set, after projection.
\end{enumerate}

This algorithm produces very reliable results. When comparing with images, we find that it catches all features that were already visually identified as cliffs, but can also isolate large boulders, outcrops, and overhangs. Table \ref{tab:cliffs} summarizes the output of our automatic detection.

\begin{table}
\caption{Output of the automatic cliff detection algorithm. A file with all results (cliff position, local gravity, height, slope, and area) is provided as supplementary material.}
\label{tab:cliffs}      
\centering          
\begin{tabular}{l r}  
\hline\hline
\multicolumn{2}{c}{cg-dlr\_spg-shap7-v1.0\_500Kfacets}\\
Facets             & 499 902\\
Slopes > 60\degree &  78 528\\
Independent cliffs &   2 633\\
Minimum height     &    13 m\\
Maximum height     &   621 m\\
\hline                  
\end{tabular}
\end{table}

\section{Results}\label{sec:results}
\subsection{Global size distribution}
\begin{figure*}
   \centering
   \includegraphics[width=\hsize]{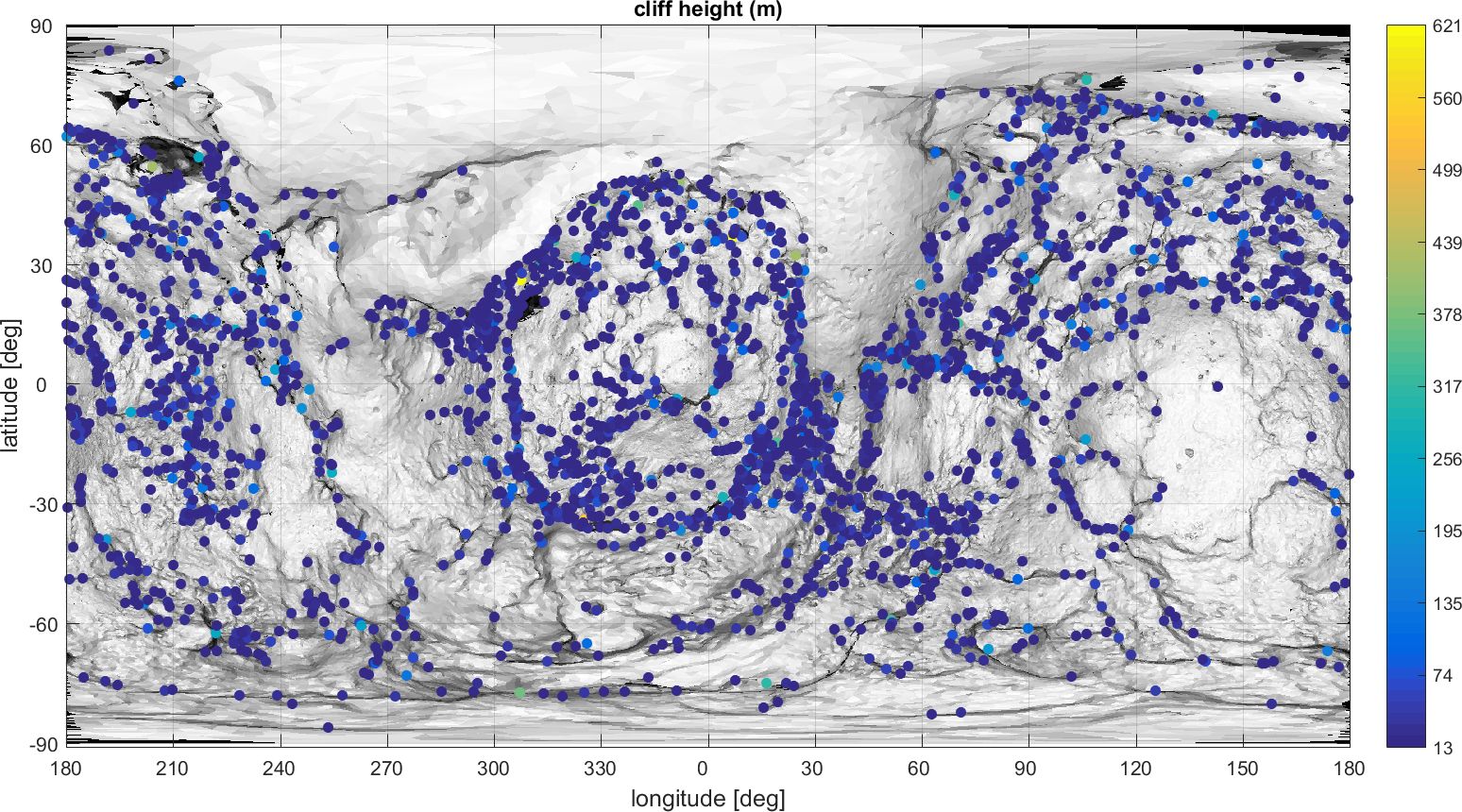}
   \caption{Cliff heights, shown as coloured dots on a shaded map of effective slope (white=flat surface, black=high slope).}
    \label{fig:map_cliffs}
\end{figure*}

Out of the 499 902 facets of the shape model, our algorithm extracted 2 633 independent "cliffs", defined as connected facets with a slope angle larger than 60\degree. Their geographic distribution is shown in Figure \ref{fig:map_cliffs}. This corresponds to 15.04\% of the total nucleus surface area. The smallest cliff detected on this shape model is 13~m high (constrained by the facet size and slope angle), while the tallest is 621~m. 

\begin{figure}
   \centering
   \includegraphics[width=\hsize]{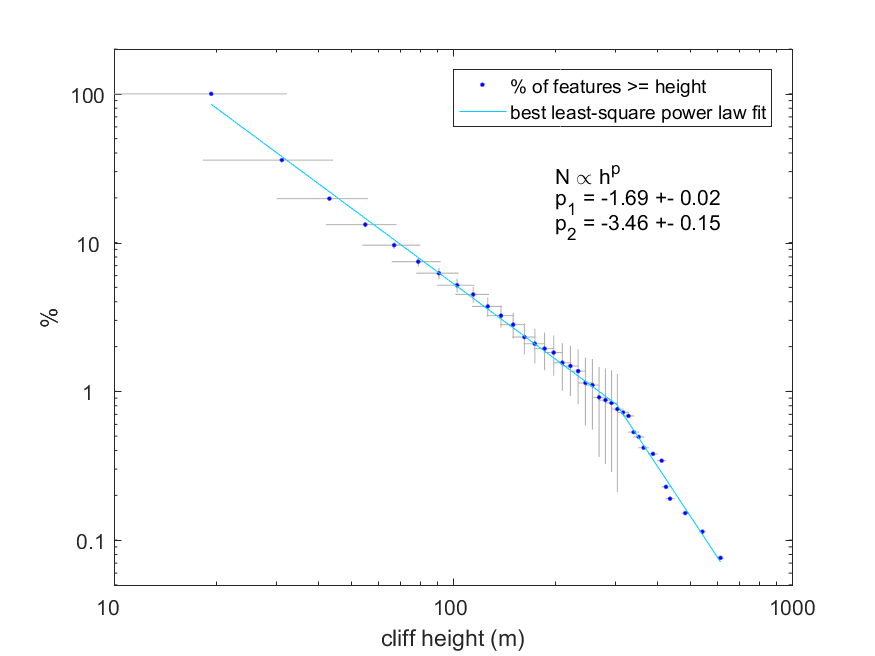}
   \caption{Cumulative distribution of cliff height on 67P/Churyumov-Gerasimenko. The vertical axis gives the percentage of cliffs taller than the height given on the horizontal axis. For instance, only 10\% of the cliffs are larger than 70~m.}
    \label{fig:cumul_plaw}
\end{figure}

The size distribution does not show any preferred height, but rather a power law, as shown in Fig. \ref{fig:cumul_plaw}. When plotting the cumulative distribution of cliff heights, we find that the lower 99.3\% of the distribution (height < 300~m) can be described with a power law index equal to $-1.69 \pm 0.02$, while the remaining 0.7\% are better represented by a power law index of $-3.46 \pm 0.15$.

\begin{figure}
   \centering
   \includegraphics[width=0.8\hsize]{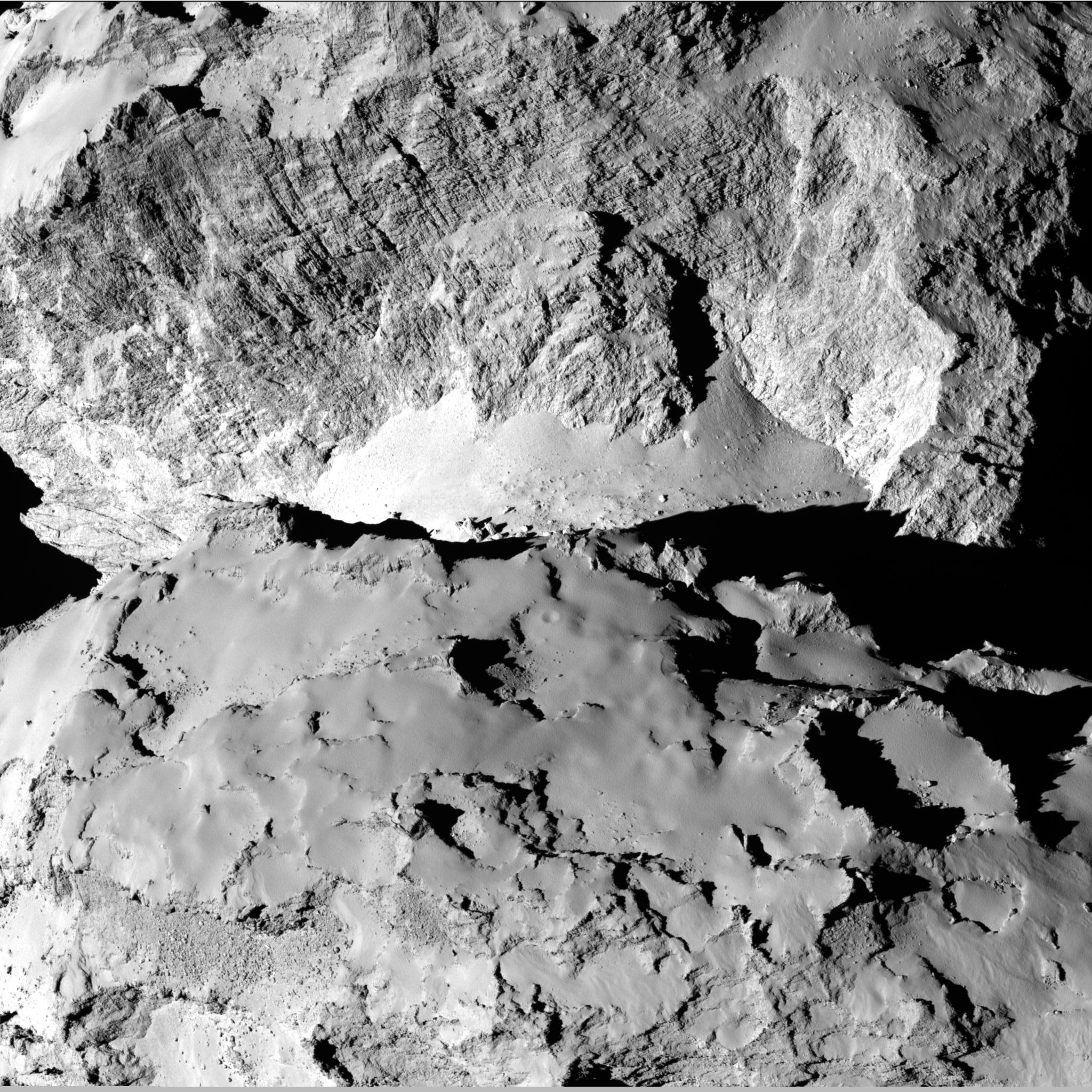}\\~\\
   \includegraphics[width=0.8\hsize]{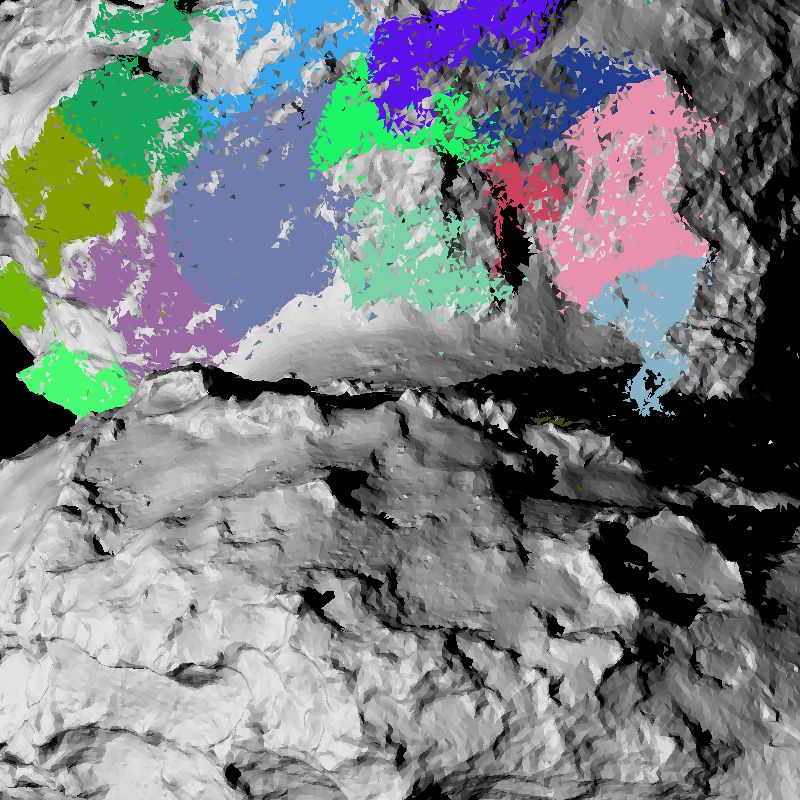}
   \caption{The largest cliffs on 67P are all located in Hathor. Top panel: OSIRIS image NAC\_2014-08-28T12.42.54.563Z\_ID30\_1397549800\_F22. Bottom panel: simulated view, colors represent the facet pertaining to cliffs taller than 250~m. Both OSIRIS image and simulated view have been rotated and aligned with the local gravity.}
    \label{fig:Hathor}
\end{figure}

The largest cliffs are mostly located in Hathor region, the area of the small lobe facing the larger component. This region oversees Hapi, the interface between both lobes of the nucleus, and has been described previously as one large cliff (\cite{thomas2015}), 900~m tall. Because its size is comparable to the small lobe itself, the gravity vector changes across the region and our automatic algorithm separates Hathor into a few distinct entities, shown in Fig. \ref{fig:Hathor}. For this reason, it is not clear whether the size distribution we observe in Hathor hints at distinct physical properties, or is rather an artefact of our definition of what a cliff is on this comet. It is interesting to note that if 67P is the result of a gentle merge between two smaller bodies as described in \cite{davidsson2016}, Hathor is effectively the former surface of the small lobe, and therefore not a cliff \textit{per se}. The significance of the power law distribution and what the different power index could mean for Hathor's material properties will be discussed in section \ref{sec:plaw}.

\subsection{Regional variations}\label{sec:regions}
\subsubsection{North vs South}
Several authors have pointed at the dichotomy between 67P's hemisphere, in terms of morphology \citep{thomas2015,elmaarry2016} or composition \citep{luspaykuti2015}. This dichotomy is largely explained by seasonal effects, as the southern hemisphere experiences significantly more erosion than the North \citep{keller2015a}. In addition to insolation, gas driven dust transport leads to a massive mantling of the Northern hemisphere \citep{lai2016} which smooths out the topography. Is this evolutionary dichotomy also present in the distribution of high slopes? 

\begin{figure}
   \centering
   \includegraphics[width=\hsize]{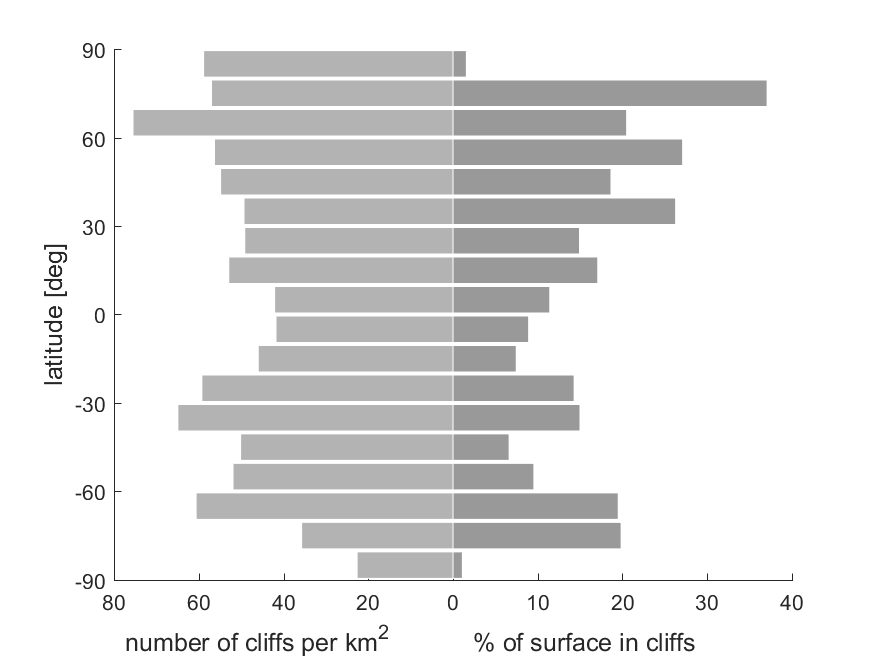}
   \caption{For each 10\degree of latitude, the left side of this plot represents the number of cliffs per square kilometre, and the right side shows how much of the area of a given latitude band is covered by cliffs. The left side can be interpreted as a measure of the roughness in the 10~m scale, while the right side is more sensitive to features in the 100~m range and beyond. Overall, this plots show that the southern hemisphere is rougher at small scales, but displays less dramatic topographic changes than the northern one.}
    \label{fig:lat_hist}
\end{figure}

Figure \ref{fig:lat_hist} shows the distribution of cliff densities (in number per $km^2$) and surface fraction as a function of the latitude. While there is no major difference in the absolute number of cliffs between the two hemispheres (50.3\% of all cliffs are in the North, 49.7\% in the South), we do observe significant variations in the local distribution:

\begin{itemize}
\item Northern cliffs are more likely to be found at higher latitudes (> 45\degree). This corresponds mainly to the Seth/Hathor regions on the big lobe, which displays some of the most dramatic topographic variations, e.g. the deep active pits presented in \cite{vincent2015b}.
\item On the contrary, southern cliffs are distributed mainly around the mid latitudes (-20\degree to -60\degree), which marks the transition area between several morphological regions \cite{elmaarry2016}. This latitude band was also identified by \cite{vincent2016b} as the preferred location for southern outbursts, many of them likely related to the sudden collapse of existing cliffs.
\item The mean density of cliffs is 5\% higher in the Southern hemisphere, but the cliffs area is proportionally larger in the North. This is effectively a quantitative measure of the surface roughness at the scale of 10-100s~m. Indeed, cliffs are more densely distributed in the South than in the dust covered North, but southern cliffs are also less high and will not tend to create large continuous walls like the ones found at high Northern latitude.
\end{itemize}

In short: the southern regions of 67P's nucleus are rougher than the northern ones at a 10~m scale, but the North is rougher at a 100~m scale. This dichotomy is a consequence of the strong seasonal differences between the two hemispheres.

\begin{figure}
   \centering
   \includegraphics[width=\hsize]{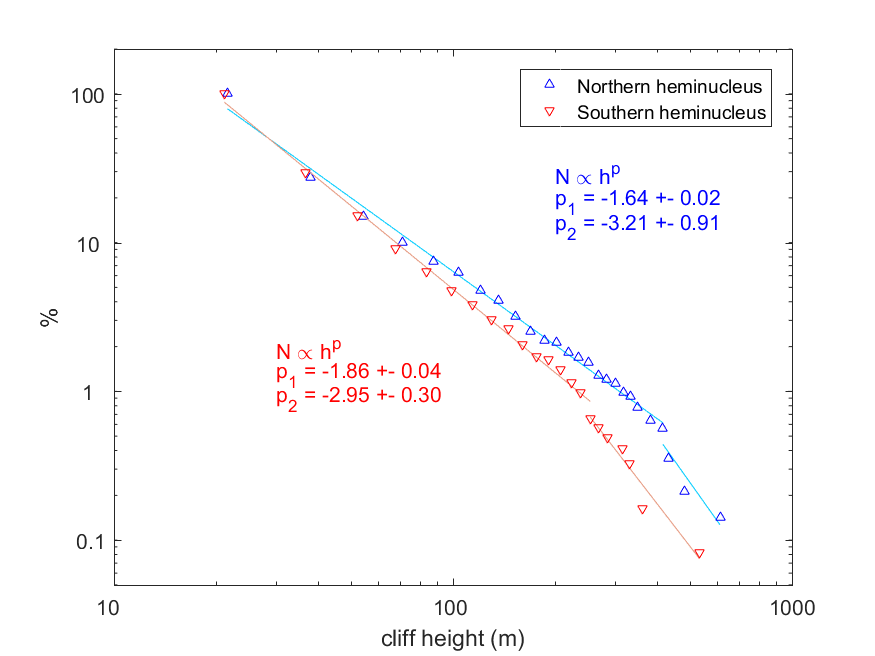}
   \caption{Cumulative size distribution of cliff height on the Northern and Southern regions of the nucleus. The southern distribution is steeper and the change of slope takes place at a lower height than on the north.}
    \label{fig:north_south}
\end{figure}

Fig. \ref{fig:north_south} shows the cumulative size distribution of cliff heights for both hemispheres. We find that the northern power index ($-1.64 \pm 0.02$) is close to the mean value of the comet, while the southern distribution shows a steeper slope ($-1.86 \pm 0.04$). Because of the different insolation patterns between both hemispheres, it is tempting to interpret this difference in power index as a signature of the surface erosional rates, or how much time the comet has spent in the inner Solar System. We develop this argument further in section \ref{sec:erosion}.

\subsubsection{Big lobe vs Small lobe}
The origin of 67P is debatable, where different publications have argued for either a primordial object \citep{davidsson2016} or a re-accreted collisional fragment \citep{rickman2015}. All authors, however, agree that 67P is very likely the result of a low-speed merger collision between two small bodies. Those objects are effectively the lobes of the comet as we see it today.

\begin{figure}
   \centering
   \includegraphics[width=0.8\hsize]{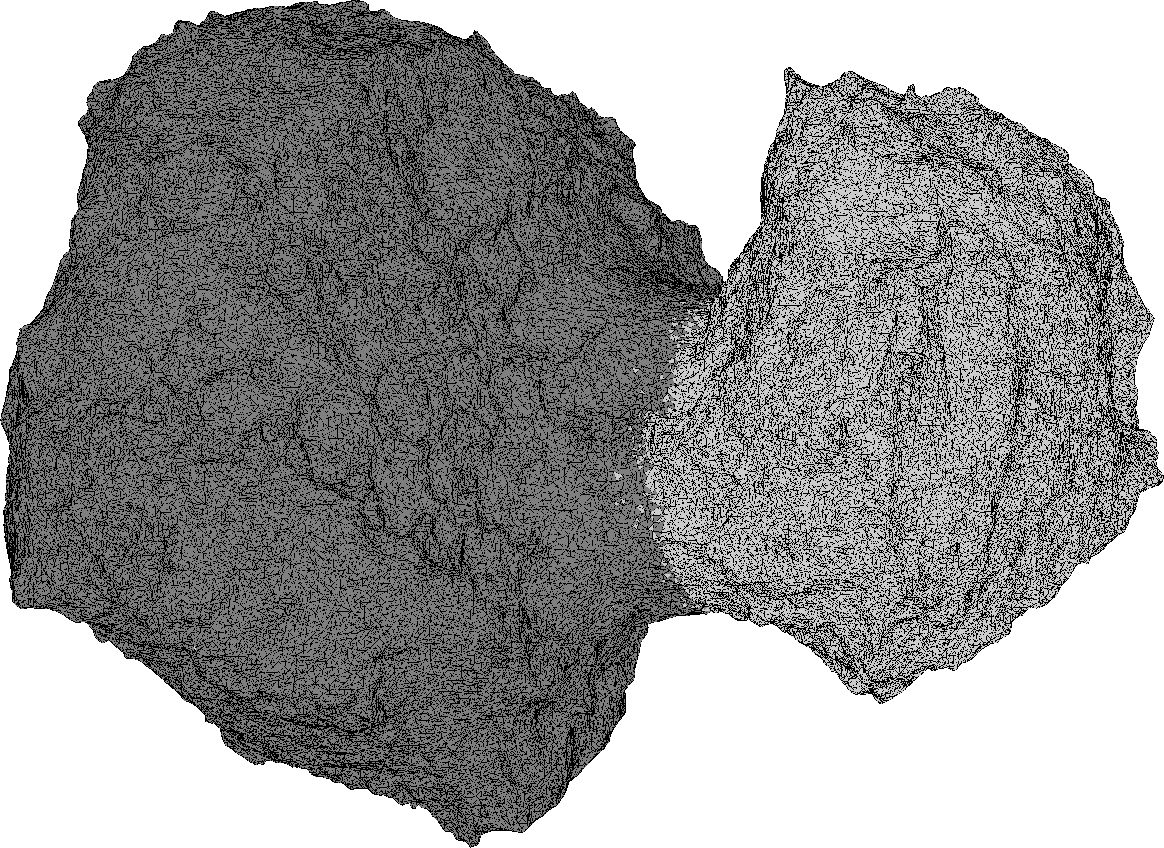}
   \caption{3D visualization of the separation between "big lobe" and "small lobe"}
    \label{fig:lobes}
\end{figure}

In our data set, the separation between the two lobes is purely geometric. \cite{preusker2015} have defined in 3D the limits of the small lobe (SL), neck region (NR) and big lobe (BL) with a set of two planes (BL-NR) and (SL-NR) which separate the shape in three entities. Vertices of the shape model belong to one component or the other depending on their position with respect to these planes. Because this definition was proposed before the Southern hemisphere was fully observed, the planes end up attributing parts of the lobes to the neck region. We correct for this by using only one separation, defined as the mean plane between the two cuts previously defined. In the Cheops-reference frame of the comet \citep{preusker2015}, a point $P[x,y,z]$ belongs to the separation plane if its coordinates satisfy the relation: 
$$1.706 x -0.846 y + 0.536 z -1.289 = 0 $$
A visualization of this separation is shown in Fig. \ref{fig:lobes}.

\begin{figure}
   \centering
   \includegraphics[width=\hsize]{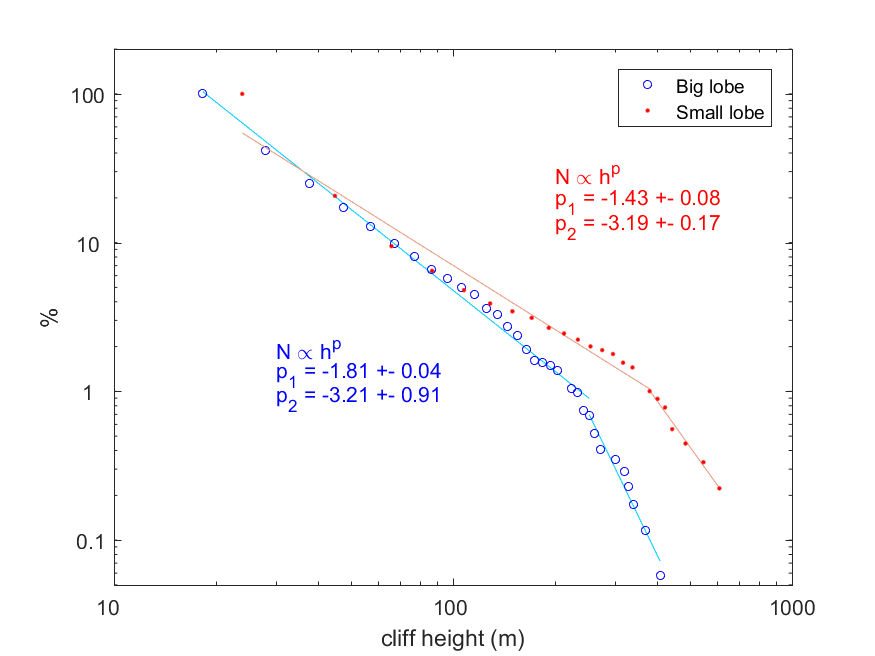}
   \caption{Cumulative size distribution of cliff height on the big lobe and small lobe. Both distributions can be approximated with a double power law. The transition from one power index to the next take place at a lower height on the big lobe.}
    \label{fig:lobes_data}
\end{figure}

~\\
We looked at the distribution of cliff heights across both lobes and summarized these results in Fig. \ref{fig:lobes_data}. We find that the big lobe follows the same trend as described earlier, with the distribution akin to a double power law (kink at 300~m). The main power index is equal to $-1.81 \pm 0.04$. The small lobe however, has a much poorer fit. The distribution can roughly be approximated with a similar set of power laws, but it is clear from Fig. \ref{fig:lobes_data} that this is not the best model. We note an excess in both the 10-20~m cliffs and the 100-200~m cliffs. This may relate to an intrinsic difference between both lobes, although it is perhaps more easily explained by a different evolution process for two main reasons:
\begin{itemize}
\item{As explained earlier, some areas in Hathor, Anuket, and Neith are formerly the original surface of the small lobe (admittedly now considerably eroded). Hence, the features pertaining to this regions that we identified as the largest cliffs now could have been flat plains when considering solely the gravity of the smaller lobe. With that in mind, it can be that those features have experienced a very different history than the smaller cliffs in other areas, and were not born as cliffs sensu stricto.}

\item{The Wosret region on the southern small lobe has a very peculiar morphology. It is extremely flat and dominated by long fractures, and devoid of any significant dust cover \citep{elmaarry2016}. Because of its location and orientation, Wosret is permanently illuminated with a Sun at zenith at perihelion. Therefore it is potentially the most eroded region of the comet, explaining why it is so flat.}
\end{itemize}

We conclude that the difference in size distribution of cliffs between the two lobe is probably not a meaningful way to assess differences in physical properties. It is, however, a good description of the different erosional history of both lobes.

\section{Discussion}\label{sec:discussion}
\subsection{Power law distribution}\label{sec:plaw}
The fact that the cliff size distribution follows a power law is not surprising, as power laws are ubiquitous in measurements of natural phenomena. Specifically in planetary science, power laws are used to best describe for instance the size distribution of craters or boulders on rocky surfaces. On 67P, we measured a cumulative power index of $-3.6 \pm 0.2$ for boulders larger than 7m \citep{pajola2015}, $-2.05 \pm 0.25$ for the diameter of circular features \citep{ip2016}, and $-2.8 \pm 0.2$ for pebbles in the Agilkia region \citep{mottola2015}. The resolution the images acquired by previous missions was not sufficient to provide an exhaustive measure of the topography, but some features (e.g. pits and boulders) have been catalogued and are listed in Table \ref{tab:all_bodies}.

\begin{table*}
\caption{Power indices (slope of the cumulative size distribution in log-log space) as measured on cometary features. The power law for circular depressions on comet 81P/Wild 2 is not provided explicitly by Basilevsky \& Keller (2006), we re-calculated it from their Fig. 10.}
\label{tab:all_bodies}      
\centering          
\begin{tabular}{l l l l}  
\hline\hline
Comet & Feature             & Power index      & Reference\\
81P   & pit diameters       & $-1.60 \pm 0.15$ & \cite{basilevsky2006} \\
9P    & pit diameters       & $-2.24 \pm 0.09$ & \cite{belton2013}\\
103P  & boulders >10m       & $-2.7  \pm 0.2 $ & \cite{pajola2016c}\\
67P   & boulders >7m        & $-3.6  \pm 0.2 $ & \cite{pajola2015}\\
67P   & pit diameters       & $-2.05 \pm 0.25$ & \cite{ip2016}\\
67P   & cliff heights       & $-1.69 \pm 0.02$ & this study\\
\hline
\end{tabular}
\end{table*}

Although it is not well understood why such distributions should be power laws, it is generally interpreted as a signature of scale invariance \citep{turcotte1986, newman2006}. Power laws distributions are alternatively found in the literature as descriptions of fractal structures and are characterized by their fractal Hausdorff dimension $d$ \citep{hausdorff1918}. Fractal dimension and power index relate to each other through the equation:
$$
d = 1 + |p_{index}|
$$
where $p_{index}$ is the power slope of the \textit{cumulative} size distribution.

\begin{figure}
   \centering
   \includegraphics[width=\hsize]{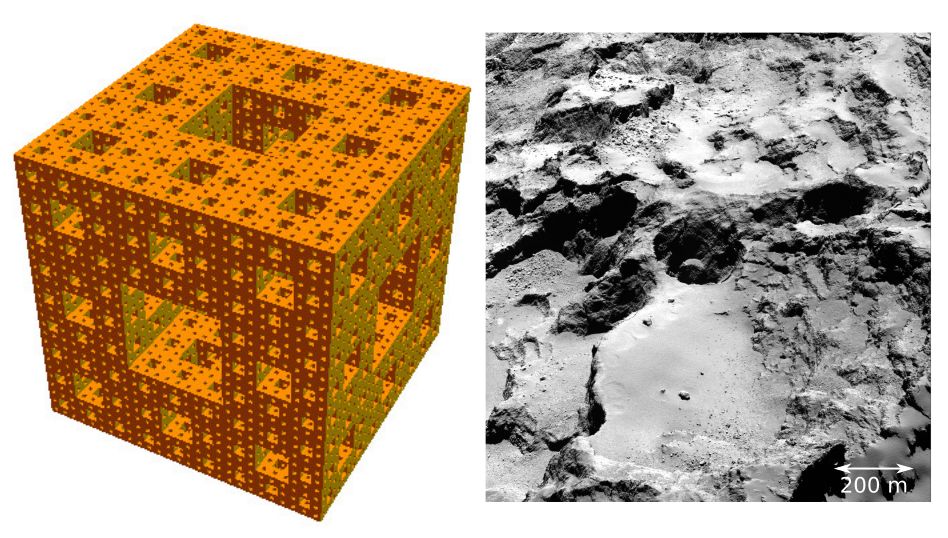}
   \caption{The \textit{Level 4 Menger sponge} (left panel) is a mathematical object with similar fractal dimension and porosity as the top $\sim 100 m$ of the comet surface, marked by large depressions, cliffs, and sharp topographic variations, as it can be seen in this OSIRIS NAC image of the Seth region (NAC\_2014-09-22T14.49.49.332Z\_ID30\_1397549400\_F22).}
    \label{fig:menger_comet}
\end{figure}

On 67P, an average $p_{index}$ of -1.69 for cliffs between 13 and 300~m is therefore equivalent to a fractal Hausdorff dimension $d = 2.69$. Hence, a pure mathematical approximation of the first 300~m of the comet could be an object such as the Level 4 Menger sponge \citep{menger1928}, which has a Hausdorff dimension $\simeq 2.73$, and 70\% porosity. Indeed, 67P's porosity is in the range 70-75\%, according to \cite{jorda2016} and \cite{paetzold2016}. This may prove useful when developing further models of the top 300 metres of the surface (Fig. \ref{fig:menger_comet}).

~\\
In terms of geophysical processes, this means that cliff formation and fragmentation tends to follow existing planes of failure which can be found at all scales. Additionally, \cite{turcotte1986} have shown that for general fragmenting processes, this fractal dimension is a measure of how efficiently the existing fractures will resist to fragmentation. Stronger material will have larger fractal dimension. In other words, the stronger the material, the steeper the power law. When applied to 67P, this observation means that as the comet crumbles, its individual fragments tend to become more resistant to subsequent failures and it may be easier for erosion processes to break down a large cliff, rather than fragmenting small boulders.

The kink in the size distribution at large heights is difficult to explain. We rule out observation bias because we certainly cannot have missed features of a few hundred meters in size after having mapped 100\% of the nucleus surface several times over more than 2 years. We see two potential explanations for the larger power index:
\begin{itemize}
\item A steeper power slope typically indicates that more erosion/fragmentation took place. Therefore, our observations could mean that cliffs larger than 300~m break up into smaller ones more efficiently than smaller features. As cliff size is a function of the ratio between gravity and cohesion, it means that cliffs larger than this limiting height might be at the edge of where gravity starts to overcome tensile strength. Hence, the amplitude of the perturbation which may trigger the collapse will be lower than for smaller cliffs. However, this effectively defines a lower limit of 2~Pa for the material cohesion, at least an order of magnitude lower than the tensile strength derived from pit collapses \citep{vincent2015a} and overhangs \citep{groussin2015a} in the same regions. Therefore it is unlikely that these large cliffs are significantly weaker than other features.

\item Rather than invoking heterogeneity in the material properties, one may instead consider insolation conditions. For example Hathor and Sobek, the two main locations for high cliffs, display very unusual erosion patterns due to their geographic position on the comet (inside large concavities). Hence, it is quite possible that erosion did not affect the cliff size distribution in these areas in the same way it modified the other regions.
\end{itemize}

We note that the kink appears at different heights depending on the regions. While this may reflect different regional history, it is more probably due to the very small number of tall cliffs over the surface (18 out of 2633) which does not allow us to constrain properly the height at which this kink occurs.

\subsection{Correlation between Surface Erosion and Power Index}\label{sec:erosion}
We suggested in section \ref{sec:regions} that the different size distributions in between hemispheres reflects the erosional history of the surface. In order to investigate this more thoroughly, we performed an orbital integration of the received insolation for the whole surface and derived an orbital erosion rate according to thermal model B in \cite{keller2015a}. More specifically, this model approximates the surface with a porous ice layer covered with a $50\mu m$ layer of small ($5\mu m$) aggregates of dust. This layer affects the heat transfer and, consequently, the sublimation of water ice. The erosion thus calculated considers only the water mass loss and is therefore a lower limit of the average erosion. 

Despite these simplifications, the results are consistent with observations of activity, erosion, and change of rotation period of the nucleus \citep{keller2015b}, and with other published models such as \cite{lai2016}.

This approach gives us a way to account for the high non-linearity of sublimation and mass loss on the comet. This is important to consider, as although the northern and southern latitudes receive about the same amount of energy per orbit, the southern insolation gets all its energy in only 8 months when close to the Sun, and therefore the erosion of the southern surface is much more dramatic than on the North.

\begin{figure}
   \centering
   \includegraphics[width=\hsize]{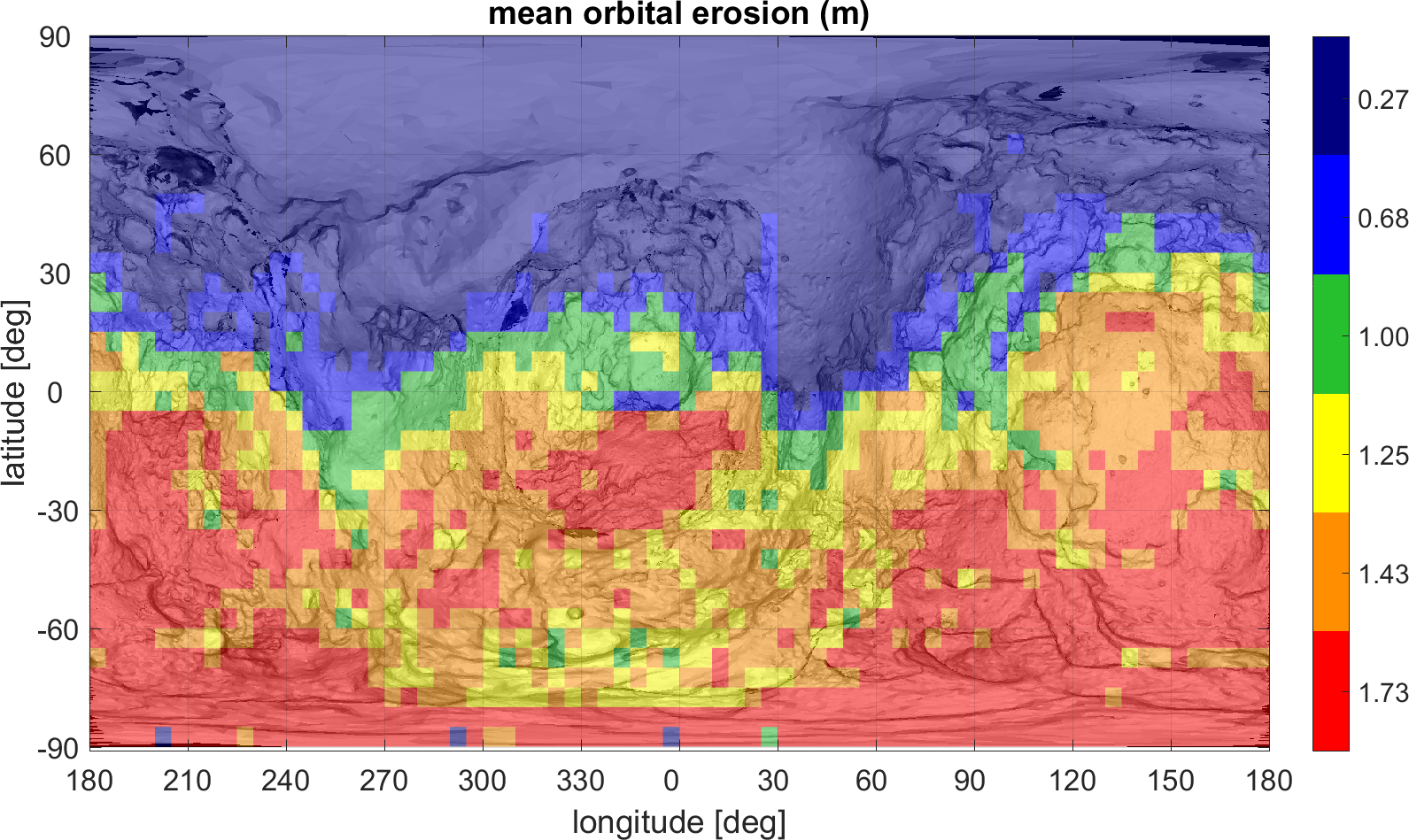}\\
   \includegraphics[width=\hsize]{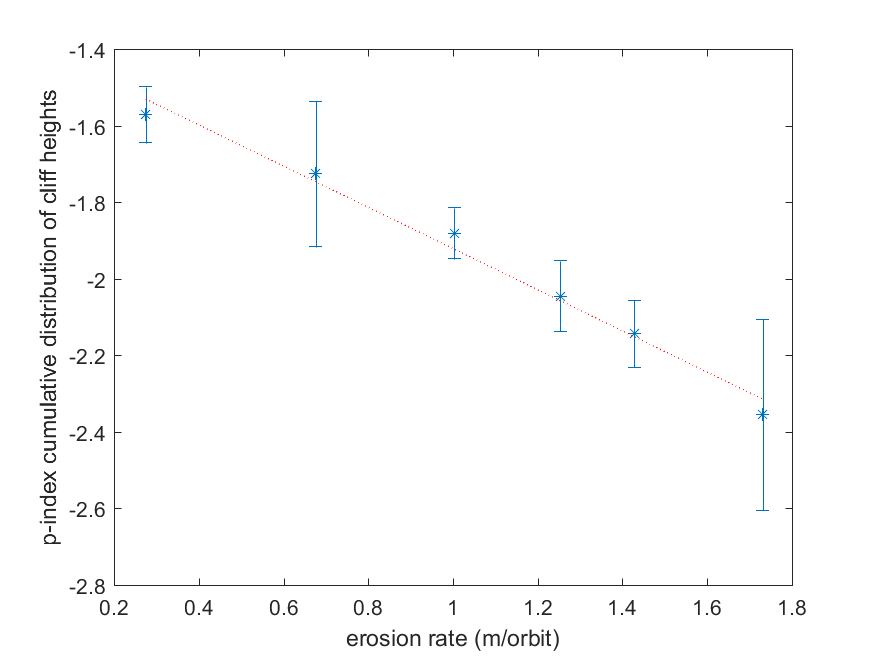}
   \caption{Top panel: Topographic map of the surface, shaded with the orbital erosion rate. Note that the equirectangular map projections makes the northernmost and southernmost regions appear larger than they are in reality. Bottom panel: power index of the cliff distribution as a function of the erosion rate. The dotted line is a linear fit to all points (correl. coeff. r=-0.993).}
    \label{fig:map_erosion}
\end{figure}

Having a model for the orbital erosion rate, we divided the nucleus surface into 6 regions with increasing erosion rates, and comparable areas and number of cliffs ($\simeq$400/region). These areas are presented in Fig. \ref{fig:map_erosion}, top panel.

For each region we calculated the power index of the cumulative distribution of cliff heights, as done before on the larger scale. Results are plotted in Figure \ref{fig:map_erosion}, bottom panel. We find a remarkable correlation (confidence 99\%) between both variables, confirming our intuition that the size distribution of cliffs is steeper for larger erosion rates.

We interpret our results as a fundamental property of erosion processes on 67P. Rather than simply losing mass, the nucleus topography is actually eroded down into ever smaller fragments that remain mostly in the regions where they were formed. The higher the erosion rate, the higher the probability to find only small cliffs. This is particularly visible when comparing for instance regions like Seth (north) which is rich in cliffs and pits with a depth >150m, with the southern Wosret that is almost completely flat. It is however important to remember that correlation does not imply causality and one cannot assume that the linear relation between erosion rate and topographic size distribution is a physical law. We can say, though, that the correlation suggest that all erosional processes (activity, thermo-mechanical stresses, gravity, ...) modify the surface in a way that is directly related to how strong and how fast the solar insolation is distributed to specific regions.

It is not clear how far this crumbling process goes as, for instance, comet pieces with a size below a few decimetres are blown away from the nucleus by activity \citep{agarwal2016}. We note, nonetheless, that the size distribution of boulders (cumul. $p_{index} = -3.6$) \citep{pajola2015,pajola2016b} and grains ejected from the comet (cumul. $p_{index} = -3~to~-~2.7$) \citep{fulle2016} is much steeper than that of the cliffs, which is compatible with our interpretation that boulders and dust are the end product of erosion.

To be exhaustive, we must also mention that although this power law evolution from shallow to steep curves seems linear for cliffs, it is not at all certain that it continues in this way for smaller blocks. Indeed, \cite{pajola2015} have shown that while most boulder size distributions follow a $p_{index}$ = -3.6, there are some areas of the nucleus with much shallower power laws ($p_{index}$ = -2, or even -1). Small objects are much more sensitive to local conditions and are certainly affected by different erosion processes than the cliffs.

\subsection{A general evolution model}
If we rewind this evolution process, can we define a primitive topographic distribution: what does a non eroded comet look like ?

We must first define what is meant by primitive in the context of cometary surfaces. In our current model of the Solar System, comets are formed beyond 30 AU and may experience a certain amount of collisions in their original environment, enhanced by the migration of giant planets. The details of this early phase are still an open question, see \cite{rickman2015} and \cite{davidsson2016} for a discussion on the potential implications of various scenarios. After this initial formation phase, comets mostly remain far from the Sun for billions of years until a favourable gravitational pull brings them back to closer heliocentric distances. Because of the low energy available, and the low density of objects at far distance from the Sun, it is likely that a cometary nucleus evolves only very little during this phase and its surface is representative of what the comet looked like shortly after formation.

Once a comet enters the inner Solar System, the situation changes dramatically, especially for Jupiter Family comets which have a small perihelion distance (e.g. 1.2 AU for 67P). We estimate that the lifetime of a comet in such orbit is in the order of a few ten thousands years, during which the surface will be completely transformed by the Solar insolation.

Reconstructing cometary orbits is notoriously difficult because of the chaotic nature of such integration (small variations in initial conditions can lead to vastly different orbits when accounting for the gravity of all planets) but the current models agree that 67P has only recently been put on its current orbit (most likely in 1959, see \cite{ip2016}). Before that time it orbited beyond the snow line, and therefore the least eroded regions on the surface are very likely to be close to their primitive state.

In the previous section we have shown how erosion affected 67P's surface: the cumulative size distribution of cliffs in the least eroded regions is characterized by a power law with $p_{index} \simeq -1.5$, while the most eroded regions have a $p_{index} \simeq -2.3$. Because of the orbital considerations expressed above, and because the most eroded areas show very little topography, we consider these two boundaries as realistic assumptions as to what the cliff height distribution should be on a very primitive and very eroded object, knowing that these can only be qualitative bounds until we have visited more comets.

From these two extreme size distributions, we propose the following evolution model: We start with a km-size body already formed; we do not consider the original accretion itself. During that formation phase, or shortly after, the topography is created with rather violent processes such as large outbursts, impacts, or self-reorganization of the nucleus constituents. These effects leave behind large topographic features on the scale of several hundred meters. The cumulative distribution of these heights is quite shallow with a power index equal or above -1.5. As the comet enters the inner Solar System, and gets eroded by activity and insolation, the topography crumbles, and the power law steepens, down to a power index equal to or lower than -2.3. Beyond that, the topography is erased and only boulders, pebbles, and dust remain. Constraining the limit at which the transition from cliffs to boulders takes place may provide important clues on the material properties. However, it requires also a precise mapping of boulder distributions as a function of erosion rate and a better understanding of the fragmentation processes, which are beyond the scope of this article.

This steepening of the power law may be partially balanced, or even counteracted by dust transport. We know from observations \citep{thomas2015,hu2017} and modelling \citep{thomas2015b,lai2016} that at least one meter of dust is deposited on regions north of +30\degree of latitude, when ejected from the southern areas at perihelion. This amounts to at least 10 metres since 67P entered the inner Solar System. This deposition would erase preferentially the smaller cliffs, and therefore make the power law shallower. Therefore the smallest power index in Figure \ref{fig:map_erosion} may not be fully representative of a primordial surface. Hence, we postulate that the original surface is more likely to look like regions of 67P that are at the same time poorly eroded and at the edge of the dust blanket (roughly in between latitudes +20\degree and +30\degree. This would correspond to the sharp cliffs/pits in Seth region on the big lobe, or the rim of the Hatmehit basin on the small lobe.

\subsection{Comparison to other bodies}
These results allow us to compare directly 67P with other comets. As Table \ref{tab:all_bodies} shows, 67P's power index for cliff heights is similar to what has been measured on 81P/Wild 2, but shallower than on 9P/Tempel 1. This is fairly consistent with observations of active pits measured by \cite{vincent2015b} which concluded that deep pits are most likely to be found in comets that have only recently entered the Inner Solar System. Smaller feature like boulders appear towards the end of this crumbling erosion, and therefore should display a steeper power law, which is observed on 67P and 103P. 

Our model suggest that 67P and 81P have encountered a similar level of erosion, while a comet like 9P, or the hyper-active 103P are more eroded. This is in agreement with our understanding of the dynamical history of these objects, both 67P and 81P for instance have entered the inner Solar System only recently \citep{brownlee2004,krolikowska2003,ip2016}. \cite{birch2017} reached a similar conclusion on the primitive state of 67P, from their analysis of several types of morphological features.

We note that one must be cautious with such comparisons as observations of other comets were acquired at much lower resolution and often describe the diameter of features rather than their height. Nonetheless, height is typically a linear function of the feature breadth (i.e. crater depth/diameter = 0.2 on most solar system bodies) and therefore should share the same power law, but this is not granted. Indeed, large boulders on 67P appear less spherical than small ones (height<diameter), and pits show at least two populations with different depth-to-diameter ratio, dominated by the eroded population \citep{vincent2015b}.

\begin{figure}
   \centering
   \includegraphics[width=\hsize]{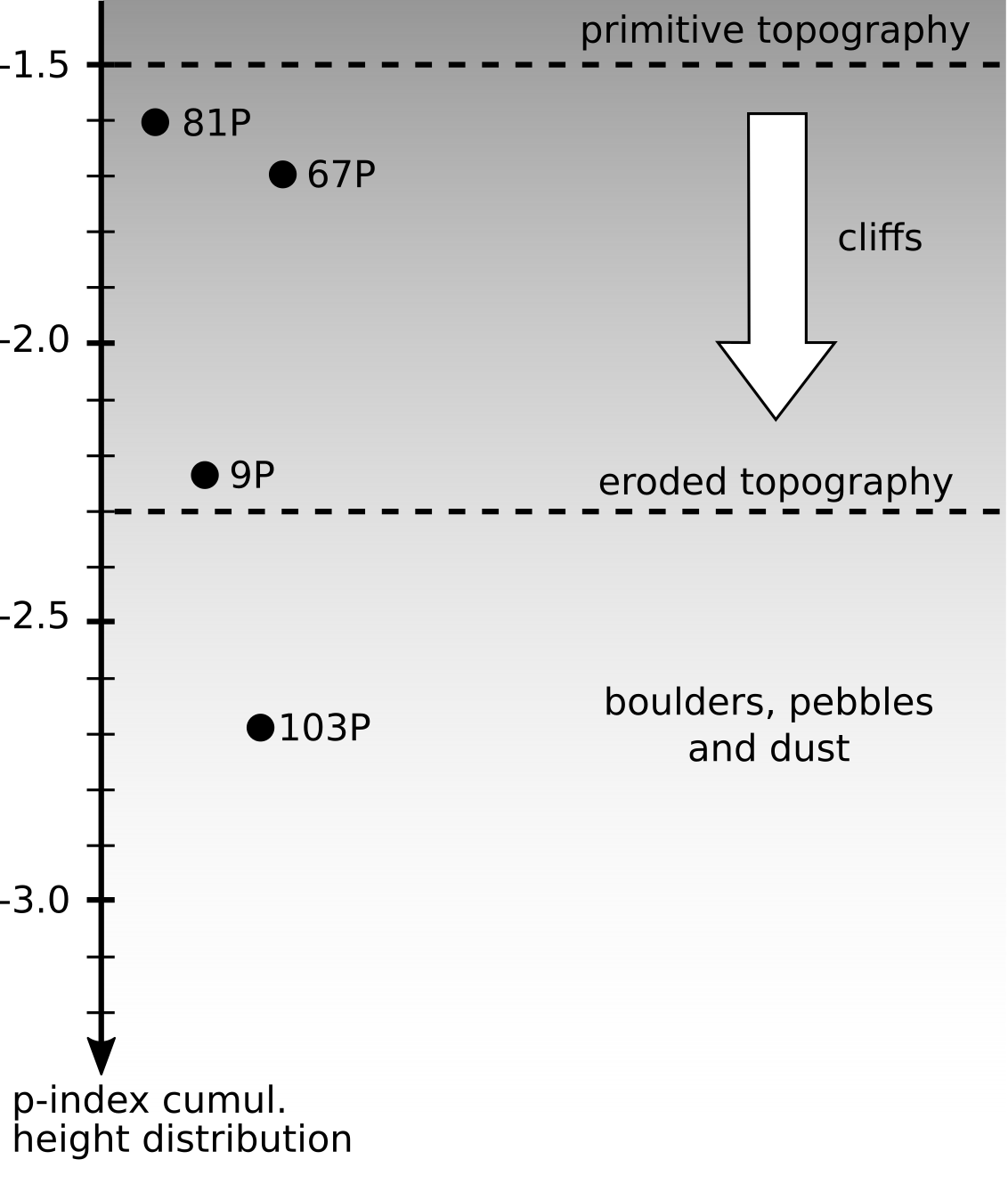}
   \caption{A model of cometary evolution. The boundaries between the different regimes are not fully constrained and should be considered qualitatively only until more comets have been characterized. The data points describe the average cumulative power index of the topographic height distribution for 4 comets and is indicative of the progression of erosion on these bodies. Because 103P is too active to sustain much topography the number given here describes the size distribution of boulders. A full list of power laws considered in this paper is given in Table \ref{tab:all_bodies}.}
    \label{fig:comet_evolution}
\end{figure}

We summarize our concept of cometary surface evolution in Figure \ref{fig:comet_evolution}, setting the boundaries for primordial and eroded comet topographies at p-indexes -1.5 and -2.3. These values are not too well constrained and require that more comets are characterized. The model is, however, qualitatively useful as it shows that a measure of the topography can provide a direct link to the level of evolution of the surface, as crater size distributions are for instance used on rocky bodies. The two boundaries can be interpreted as follows:
\begin{itemize}
\item The higher boundary (p-index $\simeq$ -1.5) defines a primordial cometary surface, shortly after formation. It reflects the events that originally shaped the topography and could provide insight on, for instance, the size and velocity distribution of small impactors in the primordial Kuiper Belt, or the intensity of early cometary outbursts. This is not an exhaustive lists of potential processes; the exploration of more comets, but also Trojans and KBOs may help us constrain this limit.
\item The lower boundary is more related to intrinsic properties of the cometary material. In essence, it describes the erosion limit at which a topographic feature cannot keep its core constituents together any more, and breaks apart into boulders, pebbles, and dust.
\end{itemize}

\section{Conclusion}
We have performed an unbiased statistical analysis of the distribution of large scale topographic features on comet 67P/Churyumov-Gerasimenko. We find that:
\begin{itemize}
\item Cliffs size distribution follow a power law with an average cumulative $p_{index}$ = $-1.69 \pm 0.02$. This slope varies from region to region, and correlates well with the orbital erosion rate of the surface. The more eroded the area, the steeper the power law.
\item This observation can be generalized to other comets. We argue that topography provides a direct measure of a comet's erosional history: primordial cometary surfaces are characterized by the presence of large cliffs, while eroded cometary surfaces are broken into smaller blocks.
\item The power law of the topography cumulative height distribution can be used as a measure of how primitive a comet nucleus is, in a similar fashion as crater counts are used to date rocky surfaces.
\item Our measurements suggest that the p-index of topographic height on a comet that has recently entered the Inner Solar system will be around -1.5. Dynamically older comets will display a larger power index, up to about -2.3.
\item Topographic features which lay outside this size distribution may be the signature of some local heterogeneity in the material properties, but most likely encountered very unusual erosion patterns due to their geographic position on the comet.
\end{itemize}

\section*{acknowledgements}
OSIRIS was built by a consortium led by the Max Planck Institut f\"ur Sonnensystemforschung, G\"ottingen, Germany, in collaboration with CISAS, University of Padova, Italy, the Laboratoire d'Astrophysique de Marseille, France, the Instituto de Astrofisica de Andalucia, CSIC, Granada, Spain, the Scientific Support Office of the European Space Agency, Noordwijk, The Netherlands, the Instituto Nacional de Tecnica Aeroespacial, Madrid, Spain, the Universidad Politecnica de Madrid, Spain, the Department of Physics and Astronomy of Uppsala University, Sweden, and the Institut f\"ur Datentechnik und Kommunikationsnetze der Technischen Universit\"at Braunschweig, Germany.

The support of the national funding agencies of Germany (DLR), France (CNES), Italy (ASI), Spain (MINECO), Sweden (SNSB), and the ESA Technical Directorate is gratefully acknowledged.

We thank the Rosetta Science Ground Segment at ESAC, the Rosetta Mission Operations Centre at ESOC and the Rosetta Project at ESTEC for their outstanding work enabling the science return of the Rosetta Mission.

This project has received funding from the European Union's Horizon 2020 research and innovation programme under grant agreement No 686709.

This research has made use of NASA's Astrophysics Data System Bibliographic Services.

The image of a Menger sponge was retrieved from \mbox{https://en.wikipedia.org/wiki/Menger\_sponge}.

Tri-dimensional visualizations in this paper are provided by the software \textit{shapeViewer} \mbox{(www.comet-toolbox.com)}.

\bibliographystyle{mnras}
\bibliography{references}

\label{lastpage}
\end{document}